# Thin Film Lithium Niobate Electro-optic Isolator Fabricated by photolithography assisted chemo-mechanical etching (PLACE)

Lang Gao,[1,2] Youting Liang,[3] Lvbin Song,[3] Difeng Yin,[1,2] Jia Qi,[3] Jinming Chen,[3] Zhaoxiang Liu,[3] Jianping Yu,[3] Jian Liu,[3] Haisu Zhang,[3,4] Zhiwei Fang,[3] Hongxin Qi,[3,4,*] and Ya Cheng,[1,3,4,5,6,7,8,*]

[1]*State Key Laboratory of High Field Laser Physics and CAS Center for Excellence in Ultra-intense Laser Science, Shanghai Institute of Optics and Fine Mechanics (SIOM), Chinese Academy of Sciences (CAS), Shanghai 201800, China*
[2]*Center of Materials Science and Optoelectronics Engineering, University of Chinese Academy of Sciences, Beijing 100049, China*
[3]*The Extreme Optoelectromechanics Laboratory (XXL), School of Physics and Electronic Science, East China Normal University, Shanghai 200241, China*
[4]*State Key Laboratory of Precision Spectroscopy, East China Normal University, Shanghai 200062, China*
[5]*Collaborative Innovation Center of Extreme Optics, Shanxi University, Taiyuan, Shanxi 030006, People's Republic of China*
[6]*Collaborative Innovation Center of Light Manipulations and Applications, Shandong Normal University, Jinan 250358, People's Republic of China*
[7]*Hefei National Laboratory, Hefei 230088, China*
[8]*Joint Research Center of Light Manipulation Science and Photonic Integrated Chip of East China Normal University and Shandong Normal University, East China Normal University, Shanghai 200241, China*
*hxqi@phy.ecnu.edu.cn
*ya.cheng@siom.ac.cn

**We report a thin-film lithium niobate electro-optic isolator fabricated by photolithography-assisted chemo-mechanical etching in this work. The device demonstrates 39.50 dB isolation when subjected to a 24 GHz microwave of 25.5 dBm on its electrodes. The measured isolation remains consistently above 30 dB within the 1510 nm to 1600 nm wavelength range. The overall device insertion loss, specifically the fiber-to-fiber insert loss, has been measured to be 2.6 dB, which is attributed to our highly efficient spot size converter and the low propagation loss observed in the fabricated waveguides.**

Nowadays, integrated photonics plays a pivotal role in developing micro- and nano-scale photonic devices[1-3]. It aims at creating compact, efficient, and cost-effective optical systems by integrating active optical components such as on-chip lasers[4, 5], amplifiers[6, 7], and photo detectors[8] with passive elements such as modulators[9, 10], beam splitters[11], filters[12], delay lines[13], and so on. To maintain the reliability and stability of the integrated photonic system, a high-performance optical isolator featuring high isolation, low insert loss, large bandwidth, small footprint, low power consumption, and compatibility with monolithic integration is urgently needed.

Previously, various types of integrated optical isolators based on magneto-optic[14, 15], acousto-optic[16, 17], electro-optic[18, 19], and optical nonlinear effects[20, 21] have been reported. Among these, electro-optic isolators show promising potential in meeting all the requirements mentioned above. However, silicon or semiconductor-based electro-optic isolators often face a trade-off between isolation and absorption loss. Recently, remarkable progress has been achieved in the fabrication of thin-film lithium niobate (TFLN) electro-optic isolators[22]. TFLN electro-optic isolators benefit from their high linear electro-optic coefficient ($\gamma_{33} = 31$ pm/V)[3, 23, 24], enabling efficient electro-optic phase modulation. Furthermore, the high refractive index ($n_o$ = 2.21 @ 1550 nm)[3, 24] of lithium niobate results in strong light confinement within its waveguides, effectively reducing absorption loss caused by the nearby electrodes. To facilitate the fabrication of photonic devices based on TFLN, photolithography-assisted chemo-mechanical etching (PLACE) technology has been successfully developed and continuously improved. This technology offers extremely low optical propagation loss, cost-effectiveness, and scalability for large-scale production[25-27].

In this study, we demonstrate a TFLN integrated optical isolator fabricated utilizing the PLACE technology. Our investigation focuses on understanding the influence of electrode length as well as input microwave power on the isolator's performance through numerical simulations. By conducting thorough theoretical analyses, we optimize the overall configuration of the electro-optic isolator. Afterwards, we successfully fabricate the device using the PLACE technology and evaluate its isolation, bandwidth response, and insertion loss performance. Our device exhibits an extremely low insert loss thanks to the efficient spot size converters and low-propagation-loss waveguides.

The principle of TFLN electro-optic isolator is depicted in Fig. 1(a). Assuming that an optical signal with amplitude $E_0$ and angular frequency $\omega_0$, denoted as $E_{in} = E_0 e^{j\omega_0 t}$, is injected into the isolator, the output signal undergoes dynamic phase modulation given by $E_{out} = E_0 e^{j\omega_0 t} e^{j\Delta\phi}$. Here, the dynamic phase induced by the applied microwave can be expressed as $\Delta\phi = \pi \frac{V_{RF}}{V_\pi} \sin(\omega_f t)$, where $V_\pi$, $V_{RF}$, and $\omega_f$ represent the half-wave voltage, peak voltage of the microwave, and angular frequency, respectively. The amplitude of the output signal can be expanded using the Bessel function of the first kind (i.e., $J_n(m), n = 0,1,...$), resulting in:

$$\text{Re}(E_{out}) = E_0 J_0(m) \cos(\omega_0 t)$$
$$+ E_0 \sum_{n=1}^{\infty} J_n(m) \cos[(\omega_0 + n\omega_f)t] \quad (1)$$
$$+ E_0 \sum_{n=1}^{\infty} (-1)^n J_n(m) \cos[(\omega_0 - n\omega_f)t]$$

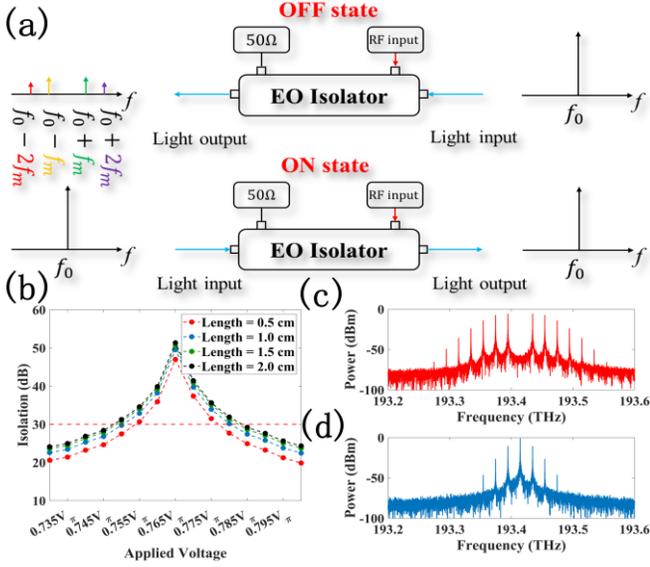

**Fig. 1.** (a) The TFLN electro-optic isolator principle schematic illustration. (b) Simulated isolation as a function of electrode lengths and microwave input powers. (c) Simulated output spectrums of the electro-optic isolator in its off-state. (d) Simulated output spectrums of the electro-optic isolator in its on-state.

Here, $m = \pi V_{RF}/V_\pi$ represents the strength of phase modulation due to the electro-optic effect caused by the microwave. When $m_{co} = \pi V_{RF}/V_\pi \approx 2.4 \approx 0.765\pi$, it leads to the condition of the 0th order Bessel function $J_0(m_{co}) = 0$, indicating complete suppression of the carrier optical signal at frequency $\omega_0$. This condition corresponds to the off-state of the isolator. However, such strong phase modulation can only be achieved when the optical signals and microwave co-propagate, i.e., travel in the same direction and satisfy the velocity matching condition for monotonic accumulation of phase modulation. In contrast, the accumulated phase modulation $m_{counter}$ typically is tiny for counter-propagating of optical signals and microwaves. Consequently, the carrier signal remains largely unmodulated due to the phase mismatch between the optical signal and microwave. This condition corresponds to the on-state of the isolator. Therefore, the device's performance can be predicted as isolation = $20 \log_{10}[J_0(m_{counter})/J_0(m_{co})]$.

We conducted numerical investigations using the Optisystem simulation software to analyze the dependence of device isolation on electrode length and microwave power at a frequency of 20 GHz. The results are presented in Fig. 1(b). Our simulations reveal that the isolation reaches its maximum when the phase modulation strength is $m_{co} = 0.765\pi$ and remains above 30 dB within the range of $m_{co} = 0.765\pi \pm 0.01\pi$. It is important to note that our simulations made assumptions of perfect velocity matching and impedance matching to focus on understanding general strategies for optimizing device performance and determining suitable device dimensions for subsequent experimental investigation. Fig. 1(b) also demonstrates that increasing the electrode length leads to higher isolation. However, beyond a length of 15 mm, the additional improvement in isolation becomes marginal. Therefore, a trade-off between device size and isolation needs to be considered. In Fig. 1(c), we present the simulated output spectra of the electro-optic isolator in its off-state when $m_{co} = 0.765\pi$. In this condition, the sideband signals are the strongest. On the other hand, Fig. 1(d) illustrates the simulated output spectra in the on-state under the same applied microwave power, indicating minimal modulation of the carrier signal, i.e., the power of carrier signal is 10 dB higher than that of the sideband signal.

The design of the TFLN integrated optical isolator is illustrated in Fig. 2(a), while Fig. 2(b) shows the fabricated device with its micrograph image. The device comprises three main components: a low-loss waveguide fabricated using PLACE technology, a pair of coplanar traveling wave electrodes with a total electrode length of 15 mm, and a pair of spot size converters (SSC) located at both ends of the device. In Fig. 2(c), we present the cross-sectional view of the device corresponding to the phase shifter section. The low-loss optical waveguide features an LN ridge waveguide with a top width of 1 μm, a bottom width of 4.8 μm, and an etch depth of 210 nm. These dimensions are obtained directly from the fabricated waveguides. The signal electrode has a width of 19.5 μm, and the gap between the electrodes placed on the two sides of the waveguide is 6.5 μm. This gap is designed to be as small as possible to achieve a low half-wave voltage while still ensuring reasonably low propagation loss for the light traveling within the waveguide.

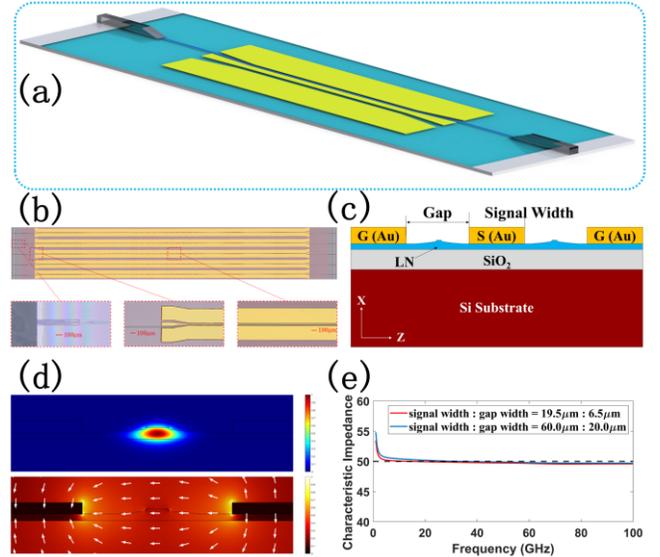

**Fig. 2.** (a) Schematic diagram of an integrated electro-optic isolator based on (TFLN). (b) Optical micrograph of the TFLN-based electro-optic isolator chip. (c) Cross-sectional structure of the phase shifter area. (d) Simulated optical field and electrical field. (e) Simulated microwave characteristic impedance of the electrodes for electrode gaps of 6.5 μm and 20 μm.

It is crucial that the two electrodes are sufficiently far away from the optical mode in the waveguide. In Fig. 2(d), we present the simulated contour plots of both optical (upper panel) and electrical (lower panel) fields. At both ends of the electrodes, taper structures are designed and fabricated to match the dimensions of the high-frequency probe, as present in Fig. 2(b). The width of the signal electrode expands from 19.5 μm to 60 μm, and the electrode gap expands from 6.5 μm to 20 μm. This design ensures good impedance matching with a 50 Ω load, as demonstrated in Fig. 2(e).

The device was fabricated on a 500-nm-thick X-cut TFLN bonded to a buried oxide ($SiO_2$) layer with a thickness of 4.7 μm. The optical waveguide fabrication utilized PLACE technology. In this process, a chromium mask was produced on the TFLN substrate using femtosecond laser surface structuring. This mask pattern was then transferred to the TFLN substrate through chemo-mechanical polishing. More detailed information can be found in Ref. [25-27]. The Ground-Signal-Ground (GSG) coplanar waveguide electrodes were prepared using femtosecond laser lithography technology followed with chemical wet etching[27, 28]. The electrode fabrication process involved several steps: (1) deposition of a 500-nm-thick gold (Au) layer using electron beam evaporation, (2) deposition of 200-nm-thick titanium (Ti) layer by magnetron sputtering, (3) patterning of the Ti layer through femtosecond laser ablation, (4) wet etching of the unprotected regions of the Au layer using aqua regia, and (5) removal of the Ti mask using sulfuric acid. These steps resulted in the successful fabrication of the desired GSG electrodes. In the final step of the device fabrication, we created spot size converters (SSC) by first forming an LN waveguide taper and then covering it with a silicon oxynitride (SiON) waveguide with a cross-sectional dimension of 4 μm × 3 μm.

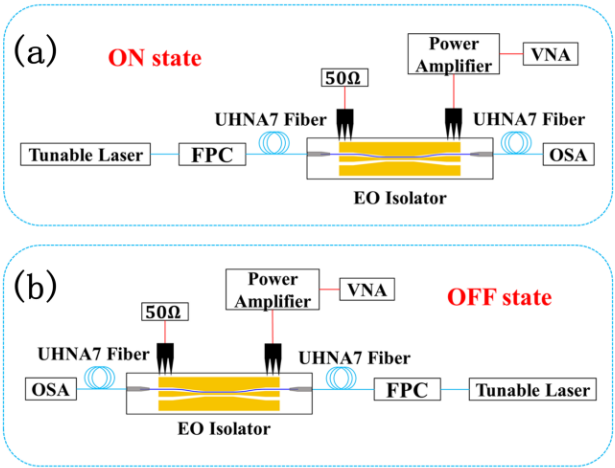

**Fig. 3.** (a) Measurement setup for the ON-state of the isolator. (b) Measurement setup for the OFF-state of the isolator.

The isolator was characterized using the experimental setup depicted in Fig. 3(a) and Fig. 3(b). Optical signals from the polarization controller were coupled into the spot size converter at one end of the waveguide, using a 3-μm-mode-field-diameter fiber. The modulated signals were then coupled back into the same type of fiber through the spot size converter. The output signals were further directed to an optical spectrum analyzer (OSA) for analysis. For microwave generation, a vector network analyzer (VNA) was employed. The VNA was connected to a microwave amplifier to amplify the microwave signal. The amplified microwave signal was injected into the microelectrodes via a Ground-Signal-Ground (GSG) probe. The opposing terminus of the electrodes was connected to a 50 Ω matching load. To measure the on-state signal, the optical signal was coupled into the waveguide from the left-hand side and counter-propagated with the microwave signal. Conversely, to switch to the off-state in the isolator, the optical signal was coupled from the right-hand side of the waveguide and co-propagated with the microwave signal.

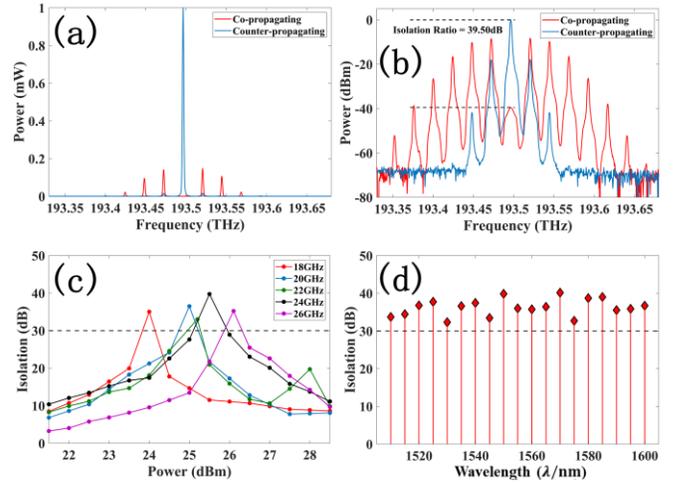

**Fig. 4.** (a) The optical spectra of co-propagating and counter-propagating conditions at a microwave frequency of 24 GHz (linear scale). (b) Optical spectrum in logarithmic scale. (c) Isolation for different microwave input powers. (d) Isolation versus optical wavelength.

Figure. 4(a) and Figure. 4(b) compare the optical spectra obtained under co-propagation and counter-propagation conditions, respectively, with the microwave frequency set at 24 GHz. In the counter-propagation (on-state) condition, the 1550 nm optical signal predominantly transmits through the waveguide with minimal loss due to sideband modulation. However, when the optical signal co-propagates with the microwave signal, a significant portion of the carrier optical signal energy is converted into sideband signals while the carrier signal is suppressed. This off-state behavior of the isolator aligns with the theoretical analysis prediction with $m = 0.765\pi$, $J_0(m) = 0$. The measured isolation reaches 39.50 dB at a microwave power of 25.5 dBm. Fig. 4(c) shows the measured isolation at different microwave input powers, which is consistent with the simulation results. Fig. 4(c) also presents the isolation at various microwave frequencies. It demonstrates that within the microwave frequency range of 18-26 GHz (which is the amplifier's gain frequency range), the measured isolation remains consistently above 30 dB. This wide bandwidth of operation is advantageous for integration with on-chip lasers, ensuring that the sideband signals can be spectrally separated from the wavelength band of gain in the lasers. It should be noted that the measured isolation is lower than the simulated isolation, primarily due to the intensity noise introduced by the microwave amplifier and the purity of the TE mode of the optical signal.

Additionally, the choice of a 15 mm electrode length is reasonable based on our simulation. The longer device length allows for efficient electro-optic modulation of the optical signal phase in the waveguide. However, this advantage will be compromised by the microwave attenuation. To assess the operation bandwidth of the isolator for different optical wavelengths, we measured the isolation at wavelengths ranging from 1510 to 1600 nm, as shown in Fig. 4(d). All measured isolations were found to be higher than 30 dB, indicating that the electro-optic isolator can achieve good isolation across a broad wavelength range. This characteristic is highly desirable for the construction of on-chip optical devices and systems. Thanks to the efficient spot size converter, we measured the fiber-to-fiber insert loss of the isolator which reached 2.6 dB.

In conclusion, we have successfully demonstrated a low insertion loss electro-optic isolator fabricated on TFLN using PLACE technique. The isolator achieves high isolation up to 39.50 dB at a microwave frequency of 24 GHz. It operates with high performance within the wavelength range of 1510 nm to 1600 nm and exhibits a fiber-to-fiber insertion loss of 2.6 dB. Our findings contribute to developing high-performance electro-optic isolators and open new possibilities for their integration into various photonic systems and applications.

**Funding.** National Key R&D Program of China (2019YFA0705000), National Natural Science Foundation of China (Grant Nos. 12192251, 12334014, 11933005, 12134001, 61991444, 12204176, 12274133), Science and Technology Commission of Shanghai Municipality (NO.21DZ1101500). Innovation Program for Quantum Science and Technology (2021ZD0301403).

**Disclosures**. The authors declare no conflict of interest.

**Data Availability.** Data underlying the results presented in this paper are not publicly available at this time but may be obtained from the authors upon reasonable request.